# Low-Cost Compact Theft-Detection System using MPU-6050 and Blynk IoT Platform

Mr. Atharva Karnik
*Dept. of Electronics and Telecommunication*
*SIES Graduate School of Technology, Nerul*
Navi Mumbai, India
atharvak@ieee.org

Ms. Diksha Adke
*Dept. of Electronics and Telecommunication*
*SIES Graduate School of Technology, Nerul*
Navi Mumbai, India
diksha@ieee.org

Prof. Pushkar Sathe
*Dept. of Electronics and Telecommunication*
*SIES Graduate School of Technology, Nerul*
Navi Mumbai, India
pushkar.sathe@siesgst.ac.in

***Abstract*— The system explained in this paper provides a compact smart surveillance system. Recent years have seen the Internet of Things (IoT) dominating in various fields of applications. With devices getting smarter and insurgence of 5G technology, the connectivity of people with devices is increasing. Smarter surveillance systems are more reliable and accessible. A gyroscope is a MEM sensor which detects angular disturbances. The principle is to detect opening or knockdown of the door physically or by a gas cutter. The system is connected to the user via Wi-Fi using ESP8266. Being a system with a low form factor, this system can be implemented on doors, shops, cars, etc. An alarm system is included in the system to alert the neighbors as well as to send a notification to the user via Blynk mobile application. The proposed system is a portable smart home solution for theft detection. The code for this system is available here: https://github.com/atharvakarnik/TheftDetectionMPU.git**

## I. INTRODUCTION

Robbery is one of the rapidly increasing threats worldwide. With advancements in technology, tools to bypass security measures are also advancing. To prevent lockpicking, advanced 7-lever, 8-lever locks were developed. Yet, these locks were brute-forced using powerful mechanical tools. In recent years, mechanical tools are getting replaced by electromechanical, chemical tools which reduced time and noise during robbery. The nation-wise data about robbery cases in Europe for year 2017 can be seen in the Fig. 1.

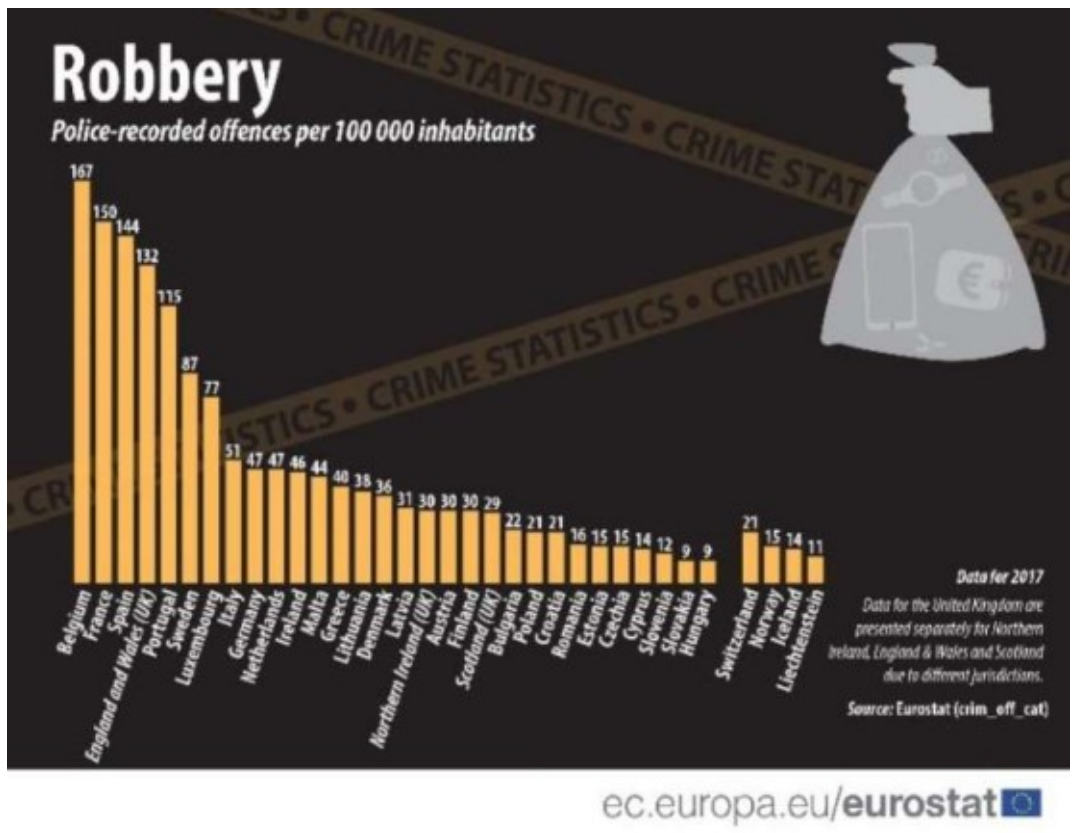

*Fig. 1. Robbery cases in Europe in year 2017*

More and more smart home management systems are developing [3][4][5][6][13][16][17] to counter this problem. In these systems, the concept of Internet of Things (IoT) is used to make devices smart and connected. The increment in the use of IoT devices can be seen in Fig. 2. Mobile phones are widely used gadget in the world. Hence there are many mobile applications such as Blynk [7], NETPIT which support Internet of Things *(IoT)*. We used Blynk mobile application to acquire updates on real time movement of the door. This system mainly focuses on making a low cost product with portability and practical application. The portability feature focuses on the device being easy to remove and implement anywhere as and when required. As per the initiative taken by Government of India to convert cities into smart cities, this system plays a crucial role of smart surveillance.

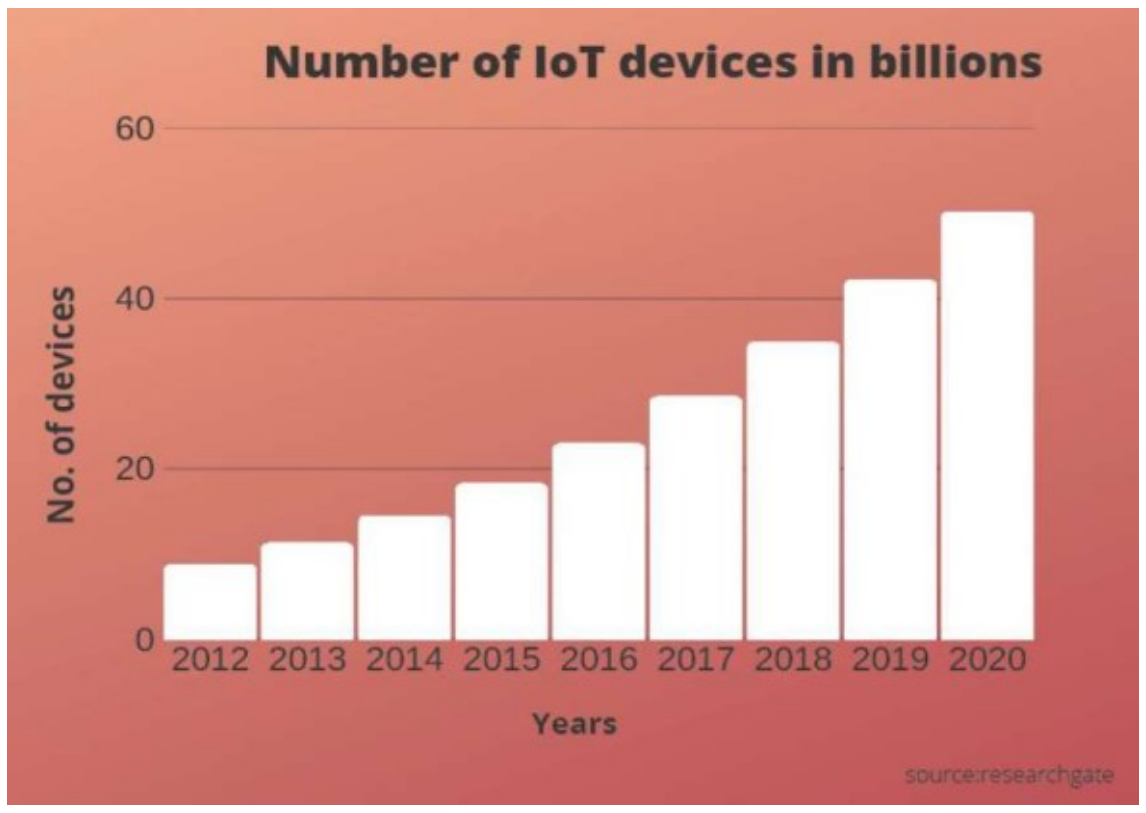

*Fig. 2. Increase in IoT devices over the years*

## II. LITURATURE SURVEY

Mamun *et al.* [1] have used GSM module, microcontroller, DTMF decoder, vibration sensor, auto dialler to detect the vehicle intrusion. If the thief tries to do any malicious activity with the vehicle, the owner will get a phone call and owner can lock the engine using a password. Engine will be started again only if the owner sends the password. As per the surveys conducted on Raspberry pi and NodeMCU [2][11][15], NodeMCU is a better choice when task is repetitive but updating code is possible in Raspberry pi only. Serikul *et al.* [3] have designed a smart capsule to monitor real time humidity in the paddy bags stored at different locations within a warehouse. Blynk mobile application was used for keeping a track of data generated from humidity sensor. When capsule is offline, supervisor will get message via blynk app.

Bohara *et al.* [4] have proposed the smart managing of electricity throughout the day which will lead in decreasing the robberies related to it. They have designed a smart home system which will do regular home chores at the fixed time every day. This will avoid peak load hours in the city. The smart home system implemented by David *et al*. [5] monitors environmental and security conditions using various sensors. The main system consist of 3 sub home automation systems and one server. The communication between systems and server is established using MTTQ protocol. It stores generated data on cloud using Thingspeak application.

Several smart surveillance systems have been designed [5][6][16] comprising of PIR sensor which detects the motion of any object, body. Saranu *et al*. developed a system [6] using a camera having an in-built CMOS image sensor and sensitive microphone which is used to record the video. This captured data will be sent to owner's mobile. The system uses a solenoid valve to secrete chloroform if any unwanted movement happens.

## III. System Specifications

### A. Hardware description

#### a. NodeMCU

The microcontroller used is ESP8266 NodeMCU, developed by ESP8266 Opensource Community. The memory and storage capacity is 128K bytes and 4M bytes respectively. The NodeMCU firmware uses the Lua scripting language. The NodeMCU has 802.11n standard wireless support (2.4 GHz), up to 72.2 Mbps

The NodeMCU works 802.11n and 802.11b networks. Therefore, it can serve as an Access Point AP, Wi-Fi station or station and AP simultaneously. The ESP8266 has full TCP/IP stack. ESP-01module allows microcontrollers to connect to a Wi-Fi network and make simple TCP/IP connections using Hayes-style commands. The processor has 16 GPIO lines, some of which are used internally to interface with other components of the SoC, like flash memory.

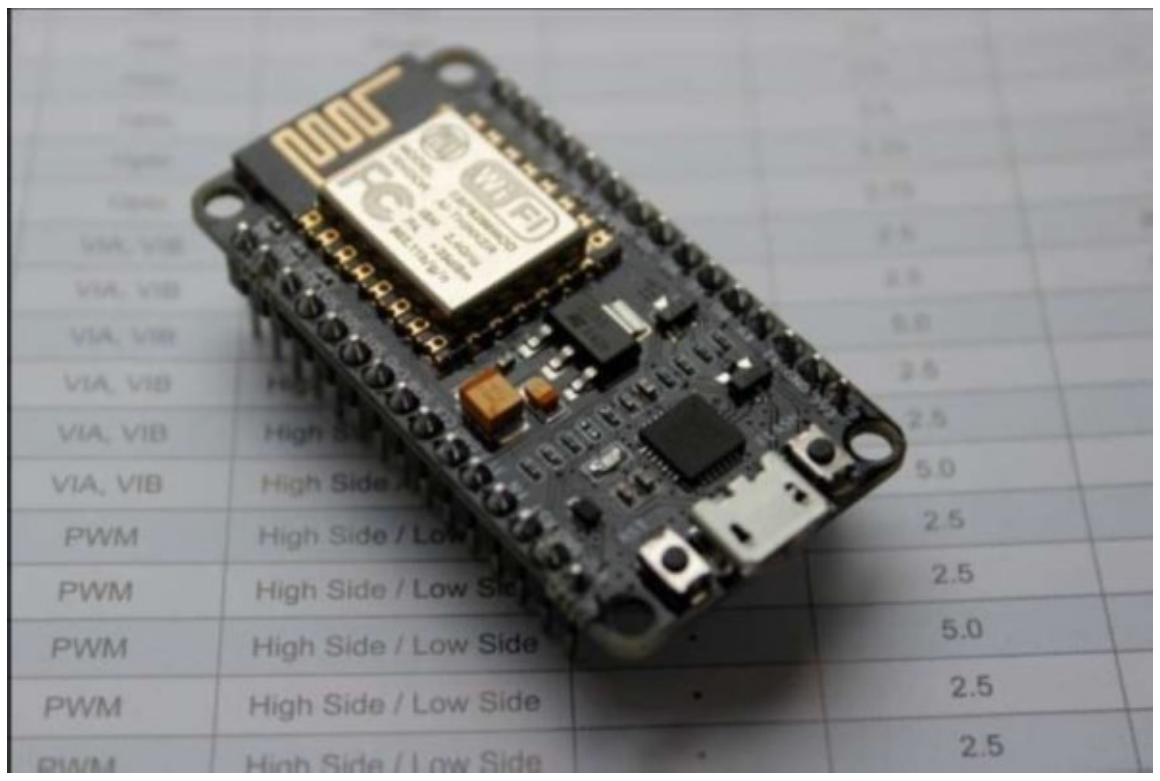

*Fig. 3. ESP8266 NodeMCU*

Since several lines are used internally within the ESP8266 SoC, 11 GPIO pins remain for GPIO purpose. Out of those 11, 2 pins are generally reserved for RX and TX in order to communicate with a host PC from which compiled object code is downloaded. Thus, there are 9 general purpose I/O pins i.e. D0 to D8. The RX, TX pins are used to communicate with other modules and devices e.g. Bluetooth communication, I2C/SPI bus, etc. The RX, TX, SD2, SD3 pins are not mostly used as GPIOs since they are used for other internal process. However, SD3 (D12) pin has the ability to respond for GPIO/PWM/interrupt like functions. The D0/GPIO16 pin can only be used as GPIO read/write, no special functions are supported on it. The block diagram of NodeMCU can be seen in the Fig. 4.

NodeMCU is a software that comes installed in ESP8266, and it uses the Lua programming language but the ESP8266 that comes with NodeMCU can also be reprogrammed via the Arduino IDE. The board has a built-in voltage regulator and the user can power up the module using the mini USB socket or the $V_{in}$ pin. Uploading code to the board is as easy as uploading code to the Arduino, there's no need for an FTDI programmer, as it comes with a USB-to-Serial converter built-in.

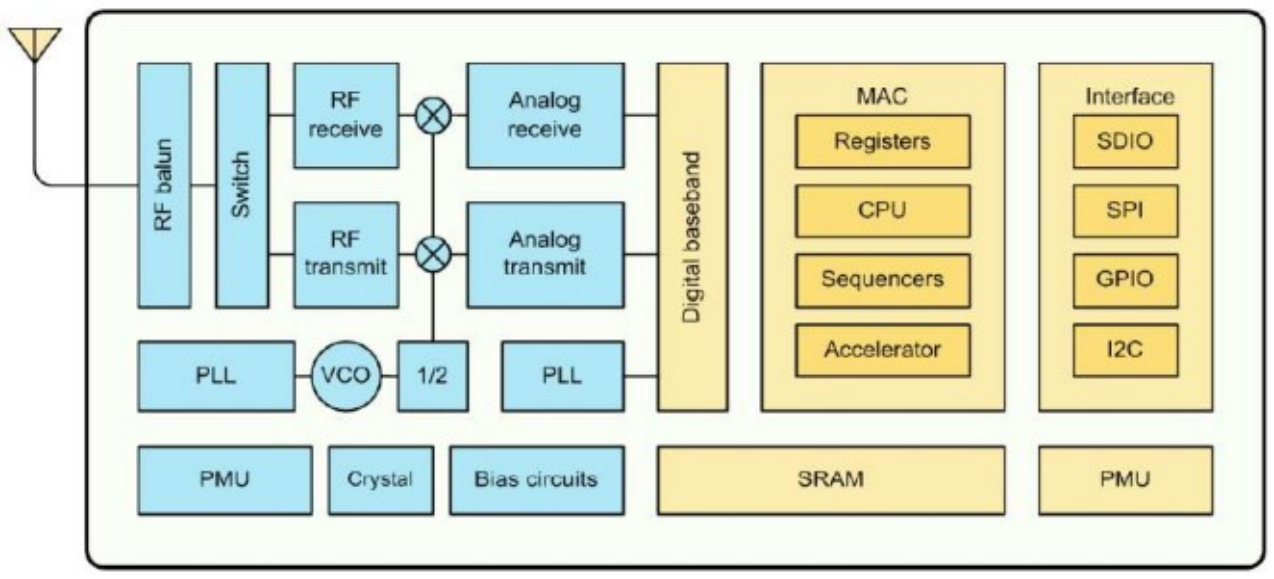

Fig. 4. Block diagram of NodeMCU

#### b. GY- 521

MPU-6050 (GY-521) is a micro-electromechanical sensor (MEMS) [8] with 3-axis accelerometer and gyroscope & temperature sensor and oscillator. The output of temperature sensor, accelerometer and gyroscope is digital with full-scale range of x-axis, y-axis, z-axis being ±250, ±500, ±1000, and ±2000°/sec. When the sensor is rotated along any axis a vibration is produced due to Coriolis effect which is detected by the MEMS. Operating voltage range of GY-521 is 2.375V-3.46V.

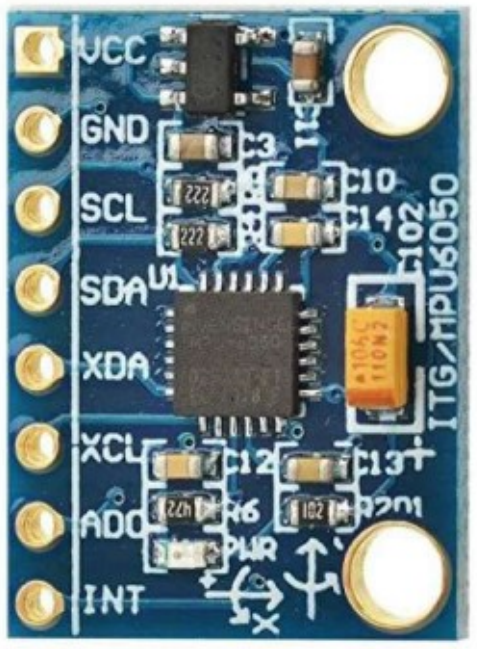

Fig. 5. GY-521 Gyroscope and Temperature sensor

GY-521 has the smallest and thinnest QFN package for portable devices: 4x4x0.9 mm. The operating current of gyroscope is 3.6mA. It has improved low-frequency noise performance. Integrated 16-bit ADCs enable simultaneous sampling of gyroscope. Operating temperature range is -40°C to +85 °C. Serial clock frequency (SCL) is 400 KHz. The Arduino and ESP8266 along with other sensors such as pressure sensor, humidity sensor, etc. can be interfaced with I2C bus. An on-chip 1024 byte FIFO buffer optimizes the power consumption of system. All the necessary processing and sensor components which are required to support motion-based use cases are provided on-chip, reducing further overall power consumption.

### c. Siren:

It is an electromechanical apparatus for generating alarm signals. Acoustic signals are used to issue forecasts, alerts and deactivations. The siren used in this solution works at 120mA current and has operating voltage range as 10.5V - 13.5V DC. Operating temperature range is -10°C - +50°C. At 12V, the output sound pressure level is 105 dB at 1 meter. Thus, it has the capability to be heard across walls for raising SOS signals.

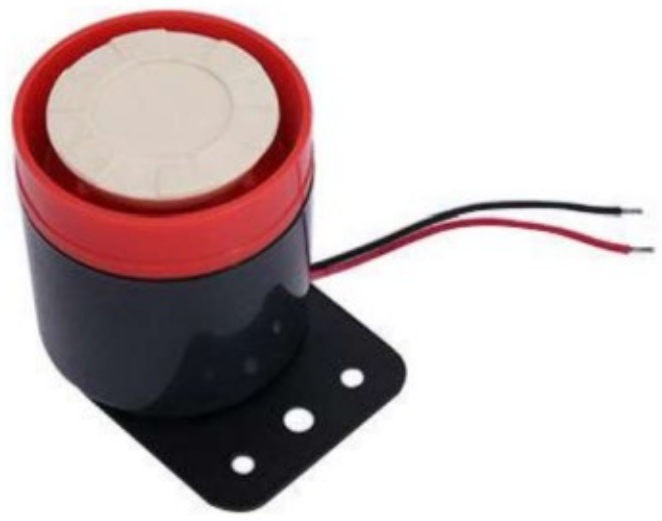

*Fig. 6. Siren*

## B. *Software Description:*

### a. Blynk

Blynk is an IoT platform [7] that supports both iOS and Android. As seen in the Fig. 7 it can compatibly work with many types of microcontrollers such as NodeMCU ESP8266, Arduino, Raspberry Pi, and ESP32 over the Internet. It consists of three major components:

i. The Blynk application, which is used to control a device and display data on widgets .
ii. The Blynk server, which is a cloud service responsible for all communications between smartphones and things.
iii. Blynk libraries, which include various widgets such as control buttons, display formats, notifications, and time management, which enable a device to send the data obtained from a sensor to be displayed on a mobile application in an effective and convenient way.

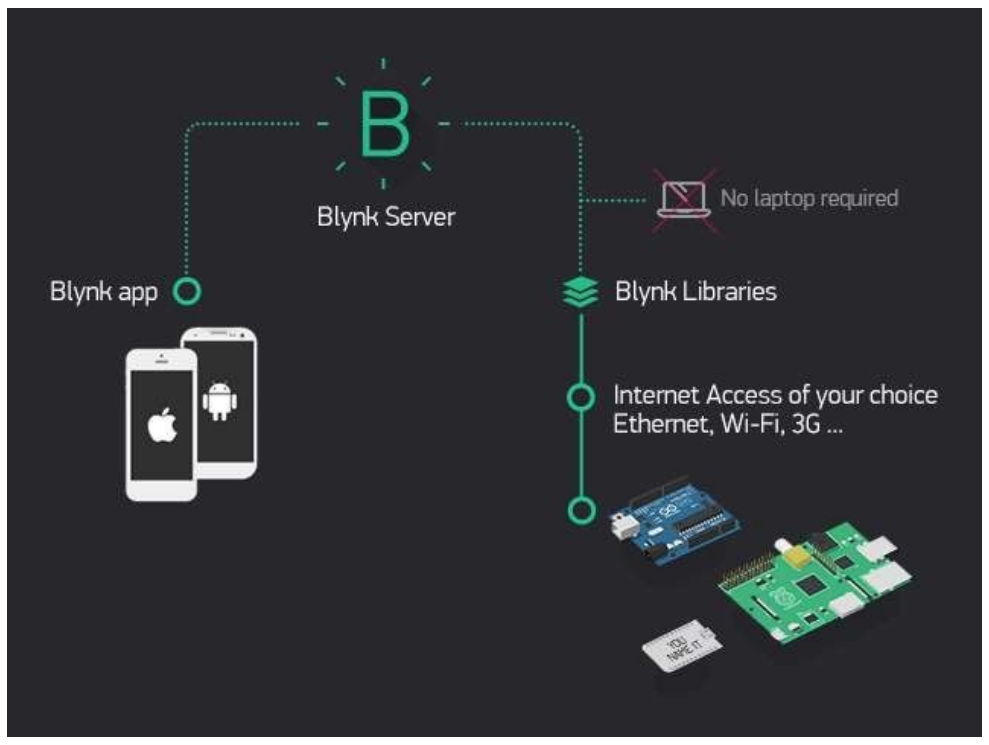

*Fig. 7. Blynk IoT application flow*

### b. Arduino IDE

The Arduino Integrated Development Environment (IDE) is an open-source application software [10] that is used to write programming codes for microcontrollers in the Arduino family. It is a cross-platform IDE which is compatible with Windows, macOS and Linux operating system. Arduino IDE can be used to execute programs for various microcontrollers, not limited to Arduino MCU. The programming language used to write instructions for the microcontroller is Embedded-C.

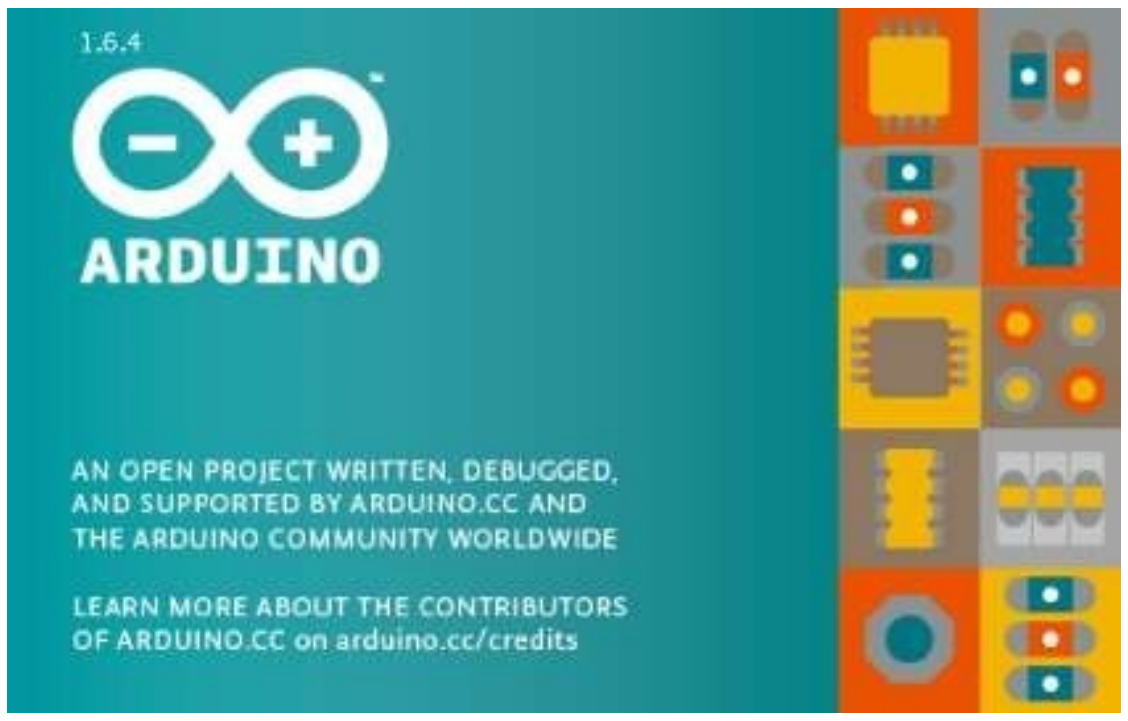

*Fig. 8. Arduino IDE*

# IV. METHODOLOGY

Till now, theft detection systems have been using laser sensors. When the beam of laser is obstructed, the alarm goes off. This kind of apparatus is however very expensive and hard to implement. It is not even feasible to implement it anywhere. Recently in 2018, Saranu *et al*. published a paper [6] about theft detection using PIR (Passive Infrared) sensor. However, it consumes much power and it is prone to raising false alarm. Also, once implemented, it is hard to relocate and reinstall it anywhere else i.e. it lacks the portability. Hence to overcome these issues, we have designed a new system with changes in the hardware. A system which is easy to install and shift anywhere as and when required.

This system uses a gyroscope sensor (MPU-6050) of GY-521 architecture. The sensor is able to detect any angular/linear acceleration. This sensor is connected with ESP8266 (NodeMCU) via I2C (Inter-Integrated Circuit) bus. Thus, when the door undergoes any physical change (movement or even breakdown), the sensor easily detects the movement. It is followed by the sensor giving excitation signal to the ESP8266. The ESP8266 module, having Wi-Fi capability can send the notification directly to the owner using Blynk mobile IoT application. It also sends signal to the burglar alarm which notifies the neighbourhood that theft is underway. The GY-521 board has an embedded ambient temperature sensor. Thus, any anomaly in temperature in case of thieves using a gas cutter can also be detected. All this functionality is achieved in a minimalistic design of very few components. This makes our design compact, portable and easy to modify. In addition to that, dedicated functionality modules *(discussed in future scope)* can also be connected to increase the overall performance of the system.

Software tools used : Arduino IDE, ESP8266 libraries, and Blynk libraries and application.

Steps to install ESP8266 and Blynk libraries on Arduino IDE :

*(Skip to step 4. If Arduino IDE is already installed, step 8. If ESP8266 library is already installed)*

1. Go to 'https://www.arduino.cc/en/main/software'
2. Choose the installation link according to your OS *(supports Windows, Macintosh, Linux)*
3. Proceed with installation
4. Start Arduino IDE and go to 'Preferences' window
5. Enter https://arduino.esp8266.com/stable/package_esp8266com_index.json' in field of 'Additional Board Manager URLs'.
6. Open Board Manager menu from Tools > Board Menu and find esp8266 platform.
7. Select the latest version from drop-down box and click Install button.
8. Now go to 'https://github.com/blynkkk/blynk-library/releases/latest' and download the .zip file. The name of downloaded file will be Blynk_Release_vXX.zip *(vXX specifies the version of library)*.
9. Unzip the file. There will be 2 folders in the archive viz. 'Libraries' and 'Tools'.
10. Copy both the folders to the Arduino IDE folder where the IDE is installed on computer and restart Arduino IDE.

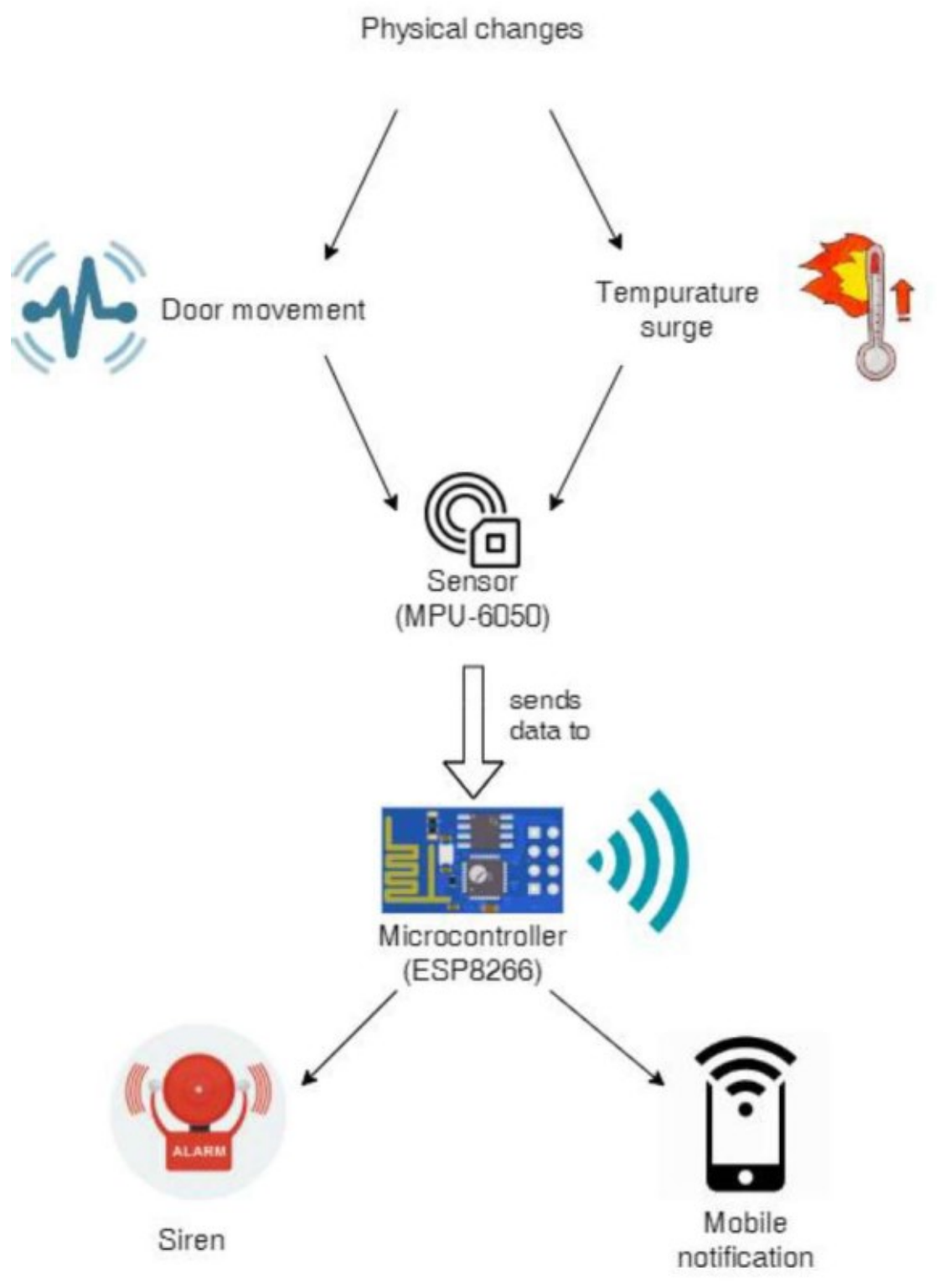

*Fig. 9. Block diagram of the system*

To proceed using Blynk library with ESP8266, we need to import '*BlynkSimpleESP8266.h*' library. As we are working with internet connection through Wi-Fi, we also need *ESP8266WiFi.h* library as well. For I2C connection of MPU-6050 and NodeMCU, we require '*Wire*' library. After compilation and uploading the code, the Serial Monitor will show 7 variables viz, Ax, Ay, Az (acceleration in x, y, z direction resp.), T (ambient temperature in °C) and Gx, Gy, Gz (angular acceleration in x, y, z direction resp.). The information flow of the system can be seen in the Fig. 9.

There have been multiple instances [18][19] where attackers tried to invade property by means of sharp objects to cut the door, powerful mechanical tools to break the lock, and in some cases, they have used gas-cutters to break-in silently. Metal doors and locks prove to be ineffective against gas cutters, thus creating vulnerability. This system presented in Fig.9. provides a reliable solution against such potential theft attacks. The system is enabled with a temperature sensor capable of measuring up to 85° C. Thus, it can easily detect rise in temperature above general ambient conditions. The data is constantly fed to microcontroller. The microcontroller, ESP8266, is connected to internet via Wi-Fi. Thus, when the sensor data crosses threshold value, notification is sent to owner via Blynk mobile application. The microcontroller also initiates the siren which has 105dB output. Thus, anyone nearby the theft location can hear the alarm and act accordingly.

## V. FLOWCHART

The Fig.10. describes the flow of tasks performed by our surveillance system. The algorithm starts by acquiring the sensor data first from the MPU-6050 sensor. The microprocessor, ESP8266 then processes the data. Two separate decisions are taken based on the threshold for temperature and threshold for angular displacement. If the values appear to be within the permissible limit, the control jumps back to reading the next data again. Any acquired value above pre-set threshold limit is translated as a malicious activity by the microcontroller. It then sends excitation signal to the siren and sends mobile notification to the owner via Blynk application.

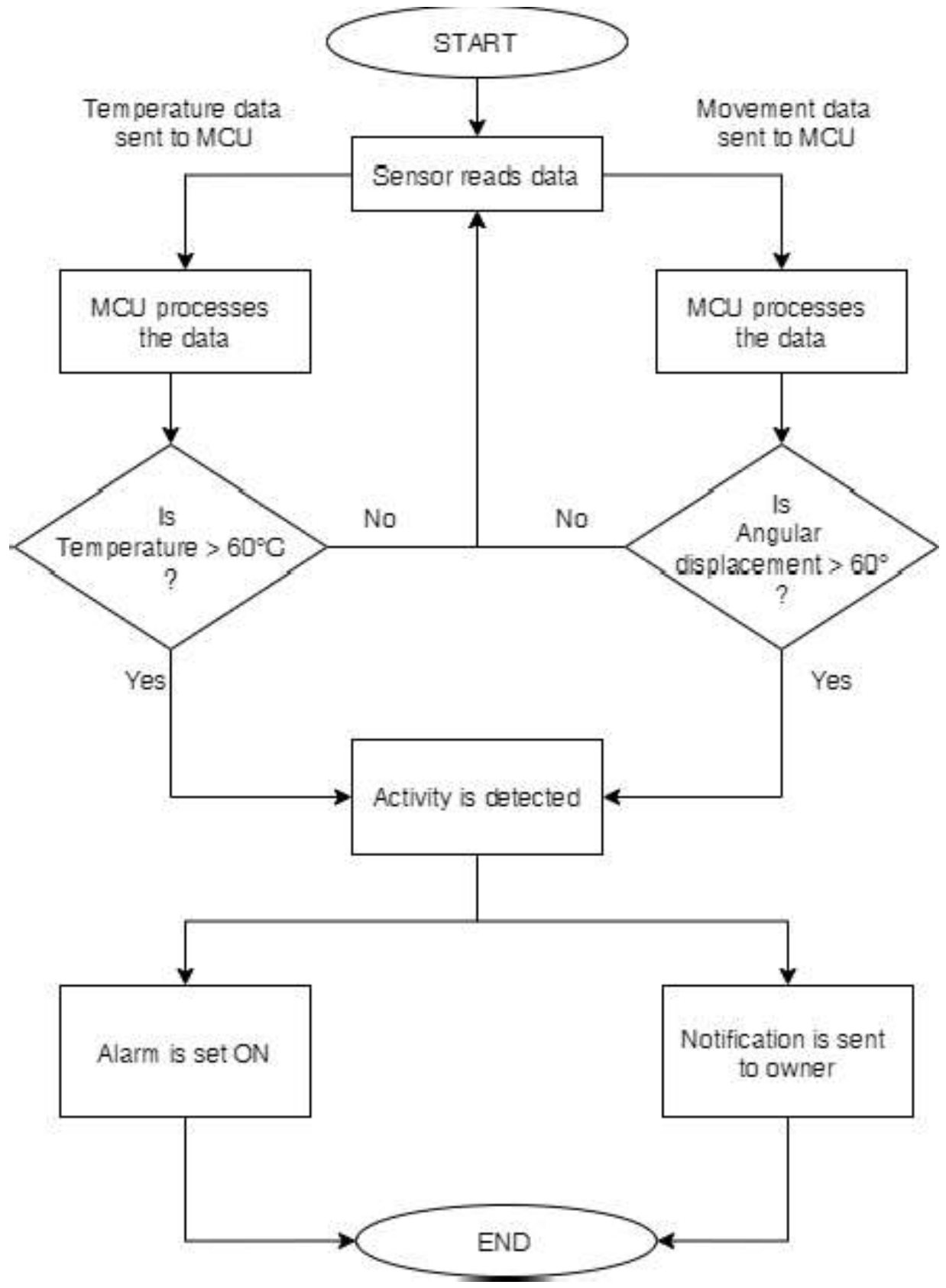

*Fig. 10. Work flow of the system*

Instead of a fixed installation like [6][16][17], we mounted the system on the door itself. Thus, any disturbances came across by the door (like opening, knock-down, cutting, etc.) can be directly picked up by the MPU-6050 sensor. Being a compact system, it can be easily implemented on the door along with microcontroller. Depending upon the orientation of mounting of Gyroscope module, vector component of motion must be chosen correctly. In our case, we chose to monitor the angular displacement in y-direction derived from the linear acceleration in x and z direction and taking its trigonometric inverse tangent function. The sensitivity of gyroscope was set to ±250°/sec. We set the threshold of angular displacement to ±60°.

This threshold was chosen experimentally to minimize false alarm due to effect of wind or other small disturbances. Thus, whenever the angular displacement in y-direction exceeded 60°, the ESP module issued excitation signal. At the moment of excitation signal, the notification was sent to user via Blynk app that ' *ALERT! The door has been opened!*' Similarly, the threshold for temperature-surge was chosen as 60°C. Above this threshold limit, a notification would get delivered to the user stating- *'ALERT! Temperature has spiked!!'*. These notifications can be seen in the Fig. 11. At the same instant, a siren included in the system would start buzzing to inform neighbours that theft is underway. The output of the system on the serial plotter can be seen from Fig. 12.

## VI. COMPARISION WITH EXISTING SYSTEM

The below Table.1. shows a brief comparison between the existing theft-detection systems using PIR sensor [6] and the gyroscope sensor used in our solution. For a fair comparison, the price of the sensor is taken from its manufacturer's website [8][9], hence it is mentioned in USD.

| Parameters | GY-521 (Gyro) | RE200B (PIR) |
|---|---|---|
| **Price** | 5.25 USD | 9.95 USD |
| **Weight** | 2.1 g | 5.87 g |
| **Form factor** | 21.2 * 16.4 * 3.3 mm | 24.03 * 32.34 * 24.66 mm |
| **Operating Voltage** | 2.375V – 3.46V | 3V–10V |
| **Temperature Measurement** | Yes | No (can only sense variations) |

*Table. 1. Comparison of existing and proposed solution*

The GY-521 sensor is nearly the half price of the RE200B PIR sensor. GY-521 is also much lighter and nearly 1/20th of RE200B in weight and cubic area respectively. It also requires lesser operating voltage (-20.8% of PIR sensor, considering minimum values). It can also measure exact ambient temperature unlike PIR sensor which has the ability to sense only the variations in temperature. Thus, being compact, power-saving and functionally efficient, GY-521 is a better choice for smart systems. The one thing it lacks is, measuring the attributes from a distance. PIR sensor is capable of getting input from wide area while GY-521 can only measure data from where it is mounted.

The Fig. 11 shows the notification received on an android mobile phone when the system detected anomalies. Fig. 12. reveals the output of GY-521 sensor on Serial Plotter of Arduino IDE.

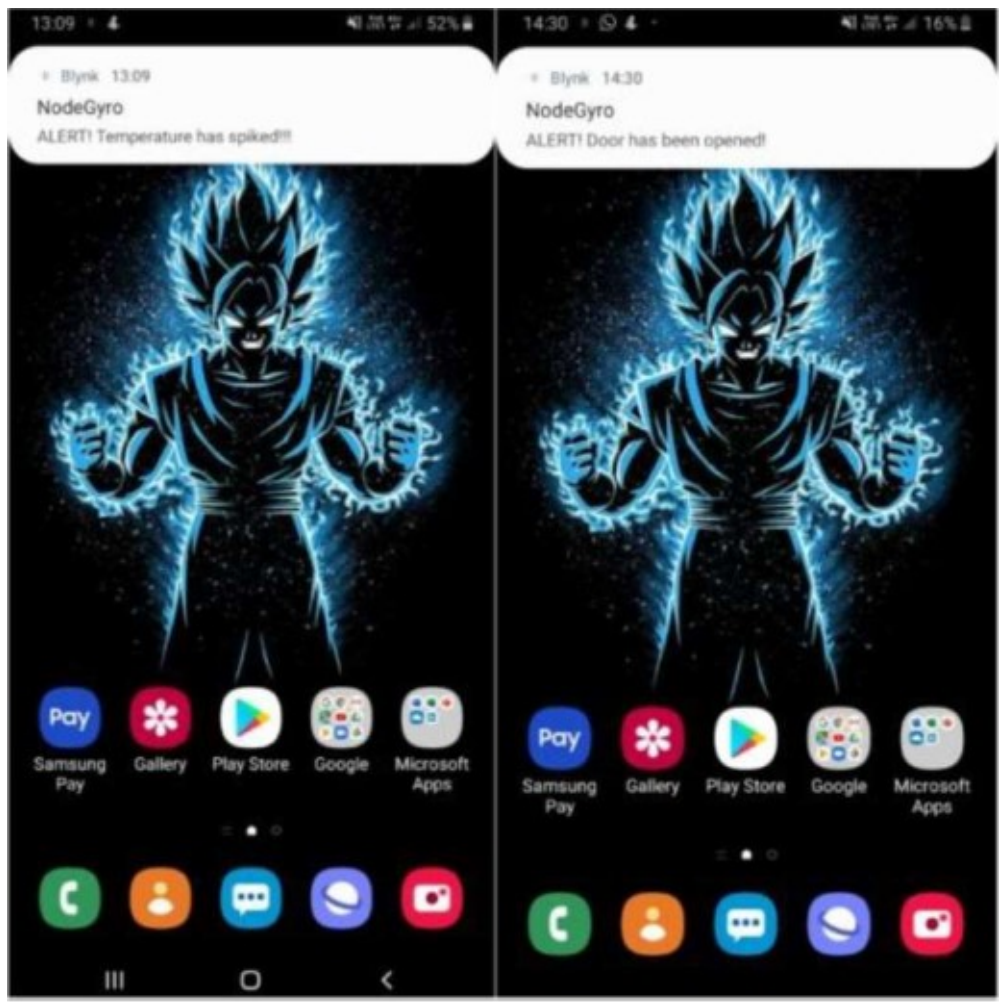

*Fig. 11. Notification received on mobile*

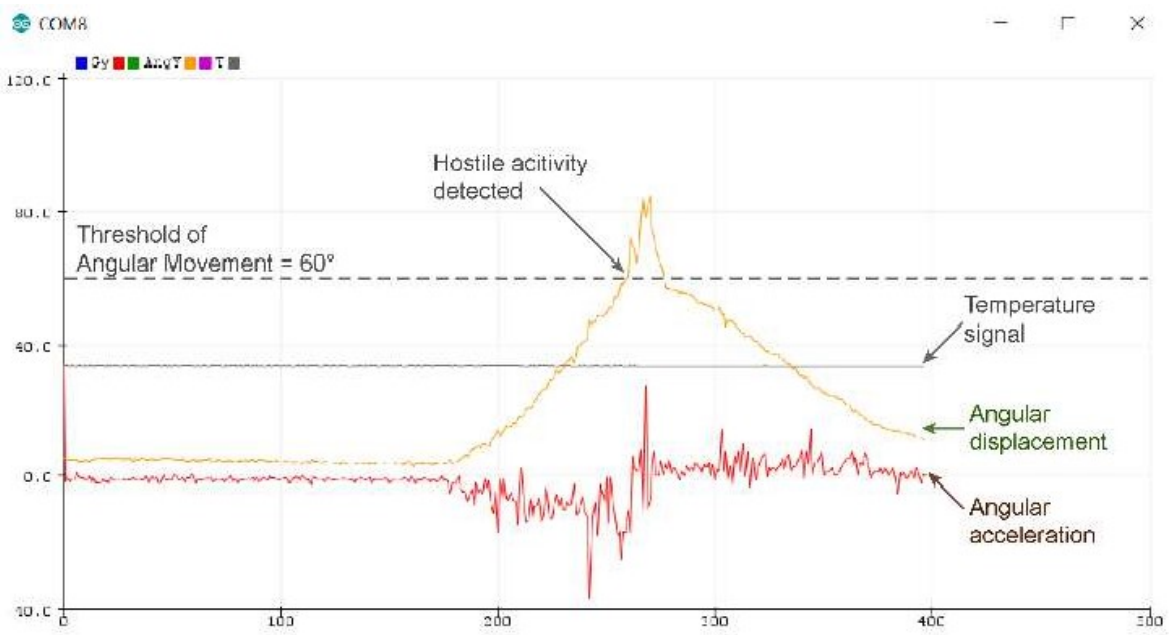

*Fig. 12. Output of the system*

## VII. FUTURE SCOPE

In some areas of world where internet connectivity is scarce, irregular, or underdeveloped, it is hard to get real-time notifications over internet. However, GSM networks can be used to notify any mischiefs. Transmission of plain text SMS is very much efficient than the transmission of codes which run android application to notify the user. For such extension, GSM modules like SIM900A can be utilized. A GSM module requires a valid SIM card and error-less code to operate. Thus, it may serve as an extension to our proposed system for functions via GSM communications. Any preventive measures or counter-attacks can also be interfaced with this system like spraying of chloroform [6] to render thieves unconscious, messaging the nearest police station about the scenario, etc.

This system may be implemented in automobile applications with suitable modifications in parameters. GPS module can be tagged along the system which may yield real-time location update of the vehicle over cloud. This may significantly increase recovery rate in the case of automobile-thefts. By using extension of GPS enabled system to our proposed system, the owner will be able to track the real-time location of the automobile.

## VIII. CONCLUSION

The project has implemented a smart home system which is able to sense any type of physical, chemical intrusion in the house. The system is not only smaller and lighter than existing solutions [3][4][5][6][13][16][17] but also provides key feature of portability i.e. it can be removed and implemented anywhere according to user's requirements. The user is able to get the notification on the mobile phone. The bulgur alarm is used to alert people in the vicinity. The power requirement of the system is very less, enabling it to function for longer duration.